\documentclass[
aps, prl,
reprint,
]{revtex4-1}
\usepackage{graphicx} 
\usepackage{subfigure} 
\usepackage{dcolumn} 
\usepackage{bm} 
\usepackage{textcomp}
\usepackage{amsmath} 
\usepackage[normalem]{ulem} 
\usepackage{xcolor} 
\usepackage{latexsym} 
\usepackage{mycommands} 

\begin{document}
\title{Chiral tunneling of topological states: towards the efficient generation of spin current using spin-momentum locking}

\author{K. M. Masum Habib}
\email{masum.habib@virginia.edu}
\author{Redwan N. Sajjad}
\author{Avik W. Ghosh}
\affiliation{Department of Electrical and Computer Engineering \\
University of Virginia, Charlottesville, VA 22904 \\
}

\begin{abstract}
    We show that the interplay between chiral tunneling and spin-momentum 
    locking of helical surface states leads to spin amplification and 
    filtering in a 3D Topological Insulator (TI). Chiral tunneling across 
    a TI $pn$ junction allows normally incident electrons to transmit, while 
    the rest are reflected with their spins flipped due to spin-momentum 
    locking. The net result is that the spin current is enhanced while the 
    dissipative charge current is simultaneously suppressed, leading to 
    an extremely large, gate tunable spin to charge current ratio ($\sim$20) 
    at the reflected end. At the transmitted end, the ratio stays close to 
    one and the electrons are completely spin polarized.
\end{abstract}


\maketitle 

Since their theoretical prediction and experimental verification in quantum 
wells and bulk crystals, Topological Insulators have been of great interest 
in condensed matter physics, even prompting their classification as a new 
state of matter\cite{qi_11}. The large spin orbit coupling in a TI leads to an 
inverted band separated by a bulk bandgap. Symmetry considerations dictate
that setting such a TI against a normal insulator (including vacuum) forces 
a band crossing at their interface, leading to gapless edge (for 2D) and 
surface (for 3D) states protected by time reversal symmetry. 
At low energies, the TI surface Hamiltonian 
$H = \vf \zuv.(\sgv\times\pv)$\cite{qi_11}
resembles the graphene Hamiltonian $H = \vf\sgv.\pv$
except that the Pauli matrices in TI represent \emph{real}-spins 
instead of \emph{pseudo}-spins in graphene.
This suggests that the chiral tunneling (the angle dependent transmission)
in a graphene $pn$ junction\cite{falko_06, young_09, sajjad_12,sajjad_13}
is expected to appear in a TI pn junction (TIPNJ) as well.
Although TIPNJs have been studied recently\cite{wu2011,takahashi2011,wang2012}, 
the implication of chiral tunneling combined with spin-momentum locking 
in spintronics has received little attention.

The energy dissipation of a spintronic device strongly depends on the 
efficiency of spin current generation.
The efficiency is measured by the spin-charge current gain
$
    \beta = \frac{2I_s/\hbar}{I_q/q}
$,
where $I_s$ and $I_q$ are the non-equilibrium spin and charge 
currents respectively. Increasing $\beta$ reduces the energy dissipation quadratically.
The gain for a regular magnetic tunnel junction is less 
than 1\cite{Datta_nanoswitch15}.
The discovery of Giant Spin Hall Effect (GSHE)\cite{Liu_GSHE_Sci12}
shows a way to achieve $\beta>1$ by augmenting the 
spin Hall angle $\theta_H$ with an additional geometrical gain\cite{Datta_CSL_APL12}.
The intrinsic gain $\theta_H$ for various metals and metal alloys has been found to vary
between 0.07-0.3\cite{mosendz2010, Liu_GSHE_Sci12, liu2011}. 
Recently, Bi$_2$Se$_3$ based TI has been reported to have `spin torque ratio'
(a quantity closely related to $\theta_H$) of 2-3.5\cite{mellnik2014} 
and has been shown to switch a soft ferromagnet at low 
temperature\cite{Fan_TI_STT_NMat14}. An oscillatory
spin polarization has also been predicted in TI using a step potential\cite{Gao_SH_TI_PRL11}.

\begin{figure}
\includegraphics[width=3in]{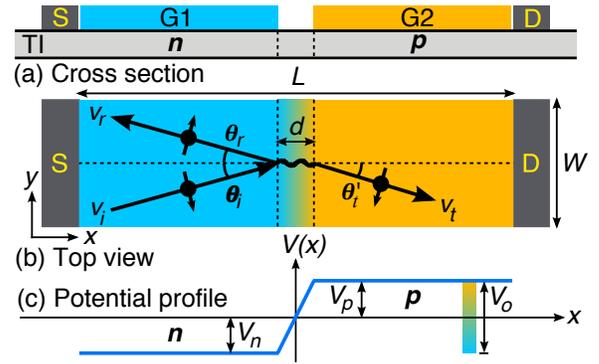}
\caption{(Color online) (a) Cross section of the TIPNJ. The source, the drain and the 
    gates are placed on the top surface of the 3D TI. The spatially separated
    gates create a graded pn junction.
    (b) Top view of the device showing the directions of incident, reflected 
    and transmitted electrons and their spins. The spin of the reflected
    wave is flipped due to spin-momentum locking which enhances the spin current
    at source.
    (c) Linear approximation of potential energy profile. 
\label{fig:device}
}
\end{figure}
In this letter, we show that the interplay between the chiral tunneling and 
spin momentum locking in TIPNJ shown in Fig. \ref{fig:device} leads to 
an extremely large, electrically tunable spin-charge current gain 
$\beta$ even without utilizing any geometric gain.
The chiral tunneling in TIPNJ only allows electrons with very small
incident angle to pass through and all other electrons are reflected back to 
the source in the same way as graphene. As a result, charge current going 
through the junction decreases. Due to spin-momentum locking, the injected 
electrons have down spin but the reflected electrons have up spin, which 
enhances the spin current at the source contact. These result in a gate 
tunable, extraordinarily large spin-charge current gain.  We show 
below that in a split-gate, symmetrically doped TIPNJ, the spin-charge current gain is,
\begin{eqnarray}
    \beta \approx \frac{1+R_{av}}{1-R_{av}} \approx \pi\sqrt{\frac{qV_od}{\hvf}}\label{eq:beta}
\end{eqnarray}
 at the source contact for small drain bias. 
Here, $R_{av}$ is the reflection 
probability averaged over all modes, $V_o$ is the built in potential of the 
TIPNJ and $d$ is the split between the gates. 
For large bias, Eq. \ref{eq:beta} can be approximated as 
$
    \beta \approx 2\sqrt{qV_od/\hvf} 
$.
In a typical TIPNJ with $d = 100$ nm, $V_o = 0.3$ V and 
$v_F=0.5\times10^6$ m/s, $\beta$ at source is $\sim$30 for small 
bias and $\sim$20 for large bias.
At drain, $\beta$ remains close to 1. We also show below 
that the $p$ region is highly spin polarized since only the small angle modes 
(with spin-$y$) exist there. 
The large $\beta$ in a TIPNJ does not require any geometrical gain and
can potentially be larger than the net gain in GSHE systems like $\beta$-Ta and W\cite{Young_APE14} that rely on the additional geometrical gain. 
In addition, it is gate tunable, meaning that we can turn its
value continuously from 1.5 to 20. 
The directions of spin and charge are parallel in TIPNJ, 
as opposed to the transverse flow in GSHE.
\begin{figure}
\includegraphics[width=3in]{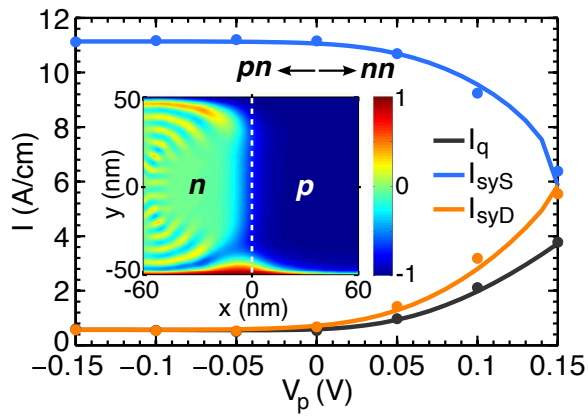}
\caption{(Color online) Charge and spin current vs. gate voltage on the $p$-side ($V_p$)
    at $V_n = 0.15$ V.
    The charge and spin currents at drain are reduced whereas 
    the spin current at source is enhanced as the device is driven from $nn$ ($V_p = 0.15$ V) 
    to $pn$ ($V_p = -0.15$ V) regime.
    The analytical (solid lines) and the 
    NEGF (circles) results are in good agreement. 
    \textbf{Inset:} Spin polarization in symmetric $pn$ regime.
	In the $p$ region only transmitted modes (spin down) exist 
	resulting in strong polarization (blue). 
	In the $n$ region, both the incident (spin down) and the reflected modes
	(spin up) exist, hence it is mostly unpolarized (green). 
\label{fig:I}
}
\end{figure}
The cross section and the top view of the model TIPNJ device are shown in 
Fig. \ref{fig:device}a and \ref{fig:device}b respectively. 
The 3D TI is assumed to be Bi$_2$Se$_3$
which has the largest bulk bandgap of $350$ meV.
The source (S) and the drain (D) contacts are placed
on the top surface of the TI slab. 
We assume that the electron conduction happens only on the top surface. This 
is a good approximation since the device is operated within the bulk bandgap 
to minimize the bulk conduction and we numerically verified that only a small 
part of the total current goes through the side walls which was also seen in
experiment\cite{Lee_surface_cond_TI_PRX14}.
The $p$ and $n$ regions are electrically doped using two external gates G1 and G2 separated by the split distance $d$. 
Such gate controlled doping of TI surface states has been
demonstrated experimentally for Bi$_2$Se$_3$\cite{Lu_gate_control_TI_PRL11}.
The device has a built-in potential $V_o = V_p + V_n$ 
distributed between the $p$ and $n$ regions as shown in Fig. \ref{fig:device}c
assuming a linear potential profile inside the
split region.
Electrons are injected from source and collected at drain by a bias voltage $V_{DS}$.

Although an equilibrium spin current exists on the TI surface,
it has no consequences for the measurable spin current\cite{Hawthorn_spin_current_PR10, Tserkovnya_12}. 
Therefore, we only considered the non-equilibrium spin current.
There has been a lot of discussions on the equilibrium spin
current in the literature\cite{Rashba_03, Tokatly_08, Sonin_07, Nikolic13}.
In this article, we choose a biasing scheme that defines 
the equilibrium state.
We connect the drain contact to the 
ground and reference the gates with respect to the ground
so that $\mu_D = 0$ and $\mu_S = qV_{DS}$ where $\mu_D$ and $\mu_S$ are the 
chemical potentials of the drain and the source contacts respectively. 
The equilibrium current, $I_{\sv_0}$ is then defined by $V_{DS}=0$ 
and $\mu_D = \mu_S = 0$. 
The non-equilibrium spin current is obtained by subtracting $I_{\sv_0}$ 
from the total spin current calculated for nonzero bias 
($\mu_D = 0$ and $\mu_S = qV_{DS}$).
A detailed description of this method is discussed in the 
Supplement.

\begin{figure}
\includegraphics[width=3in]{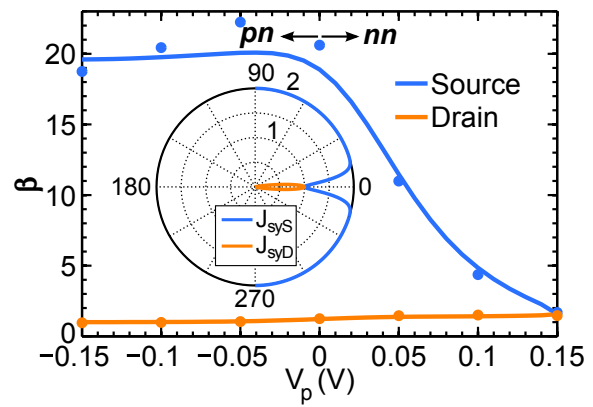}
\caption{(Color online) Spin-charge current gain $\beta$ vs. $V_{p}$ at $V_n = 0.15$ V. $\beta$ 
	increases at source as the device is driven from $nn$ to $pn$ regime. 
	The solid lines and the circles represent analytical and NEGF results respectively.
    \textbf{Inset:} Angle dependent normalized spin current densities at source and drain
    in symmetric $pn$ regime.
    Spin current at drain ($J_{syD}$) is carried by small angle 
    modes only. 
    All other modes contribute to source spin current ($J_{syS}$) twice: (1) when 
    they are injected and (2) when they are reflected since their spins 
    are flipped.
\label{fig:theta_H}
}
\end{figure}
The spin current, the charge current and the spin to charge current ratio
are shown in Figs. \ref{fig:I}-\ref{fig:theta_H} as functions of
gate bias of the $p$ region.
The solid lines were calculated 
using Eqs. \ref{eq:Jq_E_th}-\ref{eq:JsyS_E_th}
and S3 evaluated at the source and drain contacts.
The discrete points were calculated using 
the non-equilibrium Green's function (NEGF) formalism and the discretized 
k.p Hamiltonian which captures the effects of edge reflections.
Both analytical and numerical simulations were done 
for a device with length $L=120$ nm, width $W=100$ nm,
split length $d=100$ nm, drain bias $V_{DS}=0.1$ V, gate voltage $V_n=0.15$V
at room temperature. 
When the gate voltage of $p$ region $V_p = 0.15$ V, the channel is a perfect
$nn$ type with uniform potential profile. Thus, all the modes are allowed
to transmit from the source to the drain and there is no reflection. Hence,
the charge current is maximum, spin current at the source and drain are
equal and $\beta = \pi/2$ as shown in Fig. \ref{fig:theta_H}.
When the gate voltage $V_p$ is decreased to -0.15 V, the potential profile 
is no longer uniform, the channel becomes a $pn$ junction and most of the 
electrons are reflected back from the junction and therefore, charge current 
is reduced. Since the incident and reflected waves have opposite spins, the 
reflected  waves enhances the spin current at the source end and $\beta$ 
becomes large at the source contact. In the drain contact, however, only
the transmitted electrons are collected and $\beta$ remains close to 1. 
Thus, $\beta$ changes from $1.5$ to $20$ at source contact 
and remains close to $1$ at the drain when the device is driven from the $nn$ to
the $pn$ regime.
The agreement between the numerical and the analytical results shown in
Figs. \ref{fig:I} and \ref{fig:theta_H}
indicates that the physics described here is robust against the edge 
reflection at finite drain bias and room temperature.

Let us now derive Eq. \ref{eq:beta} and analyze the underlying physics. 
We start with the effective Hamiltonian for 3D TI surface states
and follow the similar procedure as 
described in Ref.\cite{sun_spin_current_PRB2005} to obtain 
the continuity equation for spin, 
$\ddp{\sv}{t} = -\nabla.\B{\hat{J}}_{\sv} + \B{\hat{J}}_\omega$.
Here, $\B{\hat{J}}_{\sv}$ is a rank 2 tensor describing the translational motion
of spin and $\B{\hat{J}}_\omega$ is a vector describing the rate of change of 
spin density due to spin precession at location $\rv$ and time $t$. The 
quantity $\B{\hat{J}}_\w$ is also referred to as spin torque\cite{sun_spin_current_PRB2005}.
Among nine elements of $\B{\hat{J}}_{\sv}$, only 
$
    \hat{J}^{x}_{sy} = -\frac{\hvf}{2}{\bf I}
$
and
$ 
    \hat{J}^{y}_{sx} = \frac{\hvf}{2}{\bf I}
$
are nonzero for TI.
The current density operator $J^{x}_{sy}$
describes spin current carried by spin-$y$ along $\hat{x}$ direction etc.
Inside the gate regions where there is no scattering, the angular term 
$\B{J}_\w$ is zero and the spin current is conserved. However, at the $pn$
junction interface, electrons are reflected which is accompanied by a change
in the spin angular momentum. As a result, inside the $pn$ junction interface,
$\B{J}_\w \neq \B{0}$ and the spin current is not conserved (see the Supplement).
At steady state, $\nabla.\B{\hat{J}}_{\sv} = \B{\hat{J}}_\omega$ 
and hence, for the two terminal device shown in Fig. \ref{fig:device}, 
the difference between the spin currents at the source and the 
drain terminal is the spin torque generated by the TIPNJ.
Similarly, we obtain the charge current density 
operators 
$
    \hat{J}^{x} = -q\vf\sgy
$
and
$
    \hat{J}^{y} = q\vf\sgx\
$
where $\hat{J}^x$ describes 
the motion of electrons moving along the $\hat{x}$ direction.
For the TIPNJ, since there is no net charge or spin transfer 
in $\hat{y}$ direction, $J^y_{sx} = 0$ and $J^y = 0$.

The wavefunction of an electron in the $n$ side ($x<-d/2$) of the TIPNJ shown
in Fig. \ref{fig:device} can be expressed as 
$
\ksi = \ket{\psi_i} + r\ket{\psi_r}
$
where $\ket{\psi_i}$ is the incident wave,  $\ket{\psi_r}$ is the reflected
wave and $r$ is the reflection coefficient. The general form of spin-momentum
locked incident wave with incident angle $\theta_i$ and energy $E$ is
$    
\ket{\psi_i} = 1/\sqrt{2A}\left(1~~-s_ii\e^{i \theta_i}\right)^{T}
    \e^{i\kv_i.\rv}
$
where $A=WL$ is the area of the device, $\kv_i$ is the wavevector with magnitude 
$k_i = \frac{|E+qV_n|}{\hvf}$ and direction $\theta_i$ and 
$s_i = \sgn(E+qV_n)$. Similarly, the reflected wave is given by 
$
    \ket{\psi_r} = 1/\sqrt{2A}\left(1~~-s_ii\e^{i\theta_r}\right)^{T}
    \e^{i\kv_r.\rv}
$
where $k_r = k_i$ and $\theta_r = \pi-\theta_i$. In the $p$ side ($x>d/2$), 
only the transmitted wave exist. Hence, the wave function of electron 
is expressed as
$\ksi = t\ket{\psi_t}$ with 
$
    \ket{\psi_t} = 1/\sqrt{2A}\left(1~~-s_ti\e^{i \theta_t}\right)^{T}
    \e^{i\kv_t.\rv}
$ 
where wavevector $k_t = \frac{|E+qV_p|}{\hvf}$, $\theta_t$ is the transmission 
angle, $t$ is the transmission coefficient and $s_t = \sgn(E+qV_p)$.
Since the potential along $\hat{y}$ is uniform, the $\hat{y}$ component of 
wavevector must be conserved throughout the device. Thus, we recover
Snell's law for TI surface state: $k_i\sin\theta_i = k_t\sin\theta_t$. It 
follows from Snell's law and the opposite helicity of conduction and valence 
bands of TI surface states that the transmission angle 
$\theta_t = \pi - \theta_t'$ for $E<-qV_p$ and $\theta_t = \theta_t'$ 
for $E>-qV_p$ where 
$
    \theta_t' = \sin^{-1}\left[\frac{E+qV_n}{E+qV_p}\sin\theta_i\right]
$.
For electrons with 
$\theta_i > \theta_c \equiv \sin^{-1}\left[\frac{E+qV_p}{E+qV_n}\right]$,
$\theta_t$ becomes complex and the electrons are reflected back to the source.

Inside the junction interface ($-d/2<x<d/2$), the wavevector varies in 
accordance with $k(x) = \frac{|E-V(x)|} {\hvf}$. 
For electrons with 
$k(x) < k_i\sin\theta_i$, the $\hat{x}$ component of $\kv(x)$ becomes 
imaginary, the wavefunctions become evanescent and the electrons are 
reflected back. 
Considering the exponential decay inside the interface and matching the 
wavefunction across an abrupt $pn$ junction, the transmission coefficient 
can be written as
$
    t = \frac{s_i\e^{i\theta_i} + s_i\e^{-i\theta_i}}{s_i\e^{-i\theta_i} + s_t\e^{i\theta_t}}\e^{-\phi}
$
where $\phi = \int\kappa(x)dx$ and
$\kappa(x) = \sqrt{k_i^2\sin^2\theta_i - k^2(x)}$ is the imaginary part of 
$\kv(x)$.

Now, let us consider an electron injected from the source at angle $\theta_i$ and 
energy $E$ is transmitted from $n$ to $p$ and collected at drain. 
The probability current density for the transmitted electron is given by
$J_{qt}(E,\theta_i) = |t|^2\bra{\psi_t}\hat{J}^{x}\ket{\psi_t}$ 
which leads to the general expression for the charge current density
\begin{eqnarray} 
    J_{q}(E,\theta_i) \equiv J_{qt} = \frac{s_tq\vf}{A}|t|^2\cos\theta_{tr}\e^{-\theta_{ti}}\e^{-\kappa_{t}L},\label{eq:Jq_E_th}
\end{eqnarray}
where $\theta_{tr}=\re{\theta_t}$, $\theta_{ti}=\im{\theta_t}$ and 
$\kappa_{t} = \im{\hat{x}.\kv_t}$. 
Similarly, the probability current density for the incident wave is
$J_{qi}(E,\theta_i) = s_iq\vf\cos\theta_i/A$.
Hence, the transmission probability is given by
$
T(E, \theta_i) 
\equiv J_{qt}/J_{qi} = \frac{cos\theta_{tr}}{\cos\theta_i}|t|^2\e^{-\theta_{ti}}\e^{-\kappa_{t}L}
$, which
is the general form of transmission probability 
in graphene $pn$ junction as presented in 
Refs. \cite{sajjad_12,falko_06} and valid for all energies in 
$nn$, $pn$ and $pp$ regime.
Similarly, the spin current density at drain is
\begin{eqnarray}
    J_{syD}(E,\theta_i) = -\frac{\hbar}{2}\frac{\vf}{A}|t|^2\e^{-2\theta_{ti}}\e^{-\kappa_{t}L},\label{eq:JsyD_E_th}
\end{eqnarray}
where the negative sign indicates that the spin current is carried by 
the down spin. 
The spin current at source has two components:
(1) the incident current
$
J_{syi}(E,\theta_i) =  -\frac{\hvf}{2A}
$
and the reflected current 
$
J_{syr}(E,\theta_i) = -\frac{\hvf}{2A}|r|^2.
$
Therefore, the total spin current density is,
\begin{eqnarray}
    J_{syS}(E,\theta_i) = -\frac{\hbar}{2}\frac{\vf}{A}(1+|r|^2)\label{eq:JsyS_E_th}
\end{eqnarray}
where $|r|^2 = 1 - |t|^2$.
Eqs. \ref{eq:Jq_E_th}-\ref{eq:JsyS_E_th} are valid for all energies 
in $nn$, $pn$ and $pp$ regimes. 
The total current is the sum of contributions from all electrons 
with positive group velocity along $\hat{x}$, weighted by the Fermi functions
and integrated over all energies as given by Eq. S3.
Unlike the incident and reflected components of charge currents,
$J_{syi}$ and $J_{syr}$ have the same sign. This is because
when a spin-up electron is reflected from the $pn$ 
junction interface, its spin is flipped due to the spin-momentum locking. 
Now, a spin-down electron going to the left has the same spin current as a spin-up electron going 
to the right. Hence, the spin currents due to the injected and the reflected 
electron add up enhancing the source spin current.

For symmetric $pn$ junction, within the barrier ($-qV_n<E<-qV_p$), the 
transmission coefficient is dominated by the exponential term  
and becomes $t\approx\e^{-\pi\hvf k_i^2d\sin^2\theta_i/2V_o}$. 
Hence, $t$ is nonzero for electrons with very small incident angle 
($\theta_i \ll \theta_c$). For these electrons, $\e^{-\theta_{ti}}\approx 1$, 
$\e^{-\kappa_{t}L}\approx 1$ and $\cos\theta_i \approx\cos\theta_{tr}$. 
Therefore, the transmission probability becomes,  
\begin{eqnarray}
    T(E, \theta_i) \approx \e^{-\pi\hvf k_i^2d\sin^2\theta_i/V_o}\label{eq:T_E_th2}
\end{eqnarray}
which has the same form as the transmission probability in graphene 
$pn$ junction\cite{sajjad_12,falko_06}.
The charge current density in symmetric $pn$ junction is then,
\begin{eqnarray}
    J_{q}(E, \theta_i) \approx q\frac{\vf}{A} [1-R(E, \theta_i)]
    \label{eq:Jq_E_th2}
\end{eqnarray}
and spin current densities at drain and source are 
\begin{eqnarray}
    J_{syD,S}(E, \theta_i) \approx -\frac{\hbar}{2}\frac{\vf}{A} [1\mp R(E, \theta_i)]		
    \label{eq:JsyD_E_th2}
\end{eqnarray}
where $-$ and $+$ signs are for $D$ and $S$ respectively, 
and $R(E,\theta_i) = 1-T(E,\theta_i)$ is the reflection probability.
Now, the spin-charge current gain can be expressed as
$
\beta(E_F) = \int d\theta 2qJ_{syS}(E_F,\theta)/\int d\theta \hbar J_q(E_F,\theta)
$
in the low bias limit. For symmetric $pn$ junction, $\beta$ at the 
source contact reduces to the first expression in Eq. \ref{eq:beta}
where 
$R_{av} = \frac{1}{\pi}\int d\theta[1-\e^{-\pi\hvf k_i^2d\sin^2\theta_i/V_o}]$ 
is the average reflection probability.
When the Fermi energy is at the middle of the barrier, $\hvf k_F = V_o/2$ and
$\beta$ is given by the second term of Eq. \ref{eq:beta}.

Eq. \ref{eq:T_E_th2} clearly shows that $T(E, \theta_i)$ is nonzero only 
for electrons with very small $\theta_i$. Hence,
only these electrons are allowed to transmit. 
For all other modes, the reflection probability 
$R(E, \theta_i)\approx 1$ and those electrons are reflected back from the 
$pn$ junction interface to the source.
Thus, only few modes with small $\theta_i$ contribute to
$J_{syD}$ and $J_q$, whereas all other modes contribute to $J_{syS}$ as shown in 
the inset of Fig. \ref{fig:theta_H}. 
This is also consistent with the spin polarization of TIPNJ shown in the inset of 
Fig. \ref{fig:I} calculated using NEGF with negligible injection from the drain.
In the $p$ side, only the transmitted waves exist and the spins of 
these electrons are aligned to $-\hat{y}$ due to the spin-momentum locking. 
Therefore, the $p$ side is highly spin polarized
as illustrated by blue.
On the other hand, in the $n$ side, both the incident 
and the reflected waves exist with spins aligned to all the directions in $x-y$ plane
leading to the unpolarized $n$ region indicated by green. This is completely
different from the uniform $nn$ or $pp$ device where the spin polarization
is $2/\pi$ throughout the channel\cite{yazyev2010,Hong_12}.
Thus, the spin polarization shown in Fig. \ref{fig:I}
is a key signature of spin filtering and amplification effect in TIPNJ,
which can be measured by spin resolved scanning tunneling 
microscopy.

One way to measure $\beta$ is to pass the spin current 
through a ferromagnetic metal (FM) by using the FM as the source contact of TIPNJ. 
The magnetization of the FM needs to be in-plane so that it does not change 
the TI bandstructure. The spin current going through the FM will exert torque 
on the FM which can be measured indirectly using spin torque ferromagnetic 
resonance technique\cite{mellnik2014} or directly by switching the magnetization 
(along $-\hat{y}$) of soft ferromagnets such as (Cr$_x$Bi$_y$Sb$_{1-x-y}$)$_2$Te$_3$
at low temperature\cite{Fan_TI_STT_NMat14}. 
Once the magnetization of the FM is switched from $-\hat{y}$ to $+\hat{y}$,
the current injection will stop (since spin up states cannot move towards right) 
and the system will reach the stable state.
%

In summary, we have shown that the chiral tunneling of helical states 
leads to an large spin-charge current gain due to the simultaneous 
amplification of spin current and suppression of charge current in a 3D TIPNJ.
The chiral tunneling allows only the near normal incident electrons to 
transmit, suppressing the charge current significantly. 
The rest of the electrons are reflected and their spins are flipped 
due to the spin-momentum locking, enhancing the spin current at the 
source end. 
The gain at drain, however, remains close to one and
the spin polarization becomes $\sim$100\%. 
Any gate controllable, helical Dirac-Fermionic $pn$ junction should exhibit
a giant spin-charge current gain which may open a new way to design spintronic devices.

This work is supported by the NRI INDEX. 
The authors acknowledge helpful discussions with Y Xie (UVa), A Naeemi
(Georgia Tech) and JU Lee (SUNY, Albany).\\
%
%
%
\begin{center}
{\LARGE \textsc{Supplemental}}
\end{center}

\setcounter{equation}{0}
\setcounter{figure}{0}

\renewcommand{\theequation}{S\arabic{equation}}
\renewcommand{\thefigure}{S\arabic{figure}}

\section{NEGF and k.p Method}
The discrete points in Figs. 2-3 (of main text) were calculated using 
the non-equilibrium Green's function (NEGF) formalism and the discretized 
k.p Hamiltonian, which captures the effects of edge reflections. Here we 
describe the calculation method.

The low energy effective Hamiltonian to describe the surface states of TI 
has been shown to be\cite{zhang_09}
\begin{eqnarray*}
H = \vf \zuv.(\sgv\times\pv)
\end{eqnarray*}
where $\pv$ is the momentum, $\sgv = (\sgx, \sgy)$ are the Pauli matrices,
and $\vf$ is the Fermi velocity of electron on the TI surface state.
To avoid the well known fermion doubling problem\cite{Stacy_82,Susskind_77} 
on discrete lattice, we added a $\sgz$ term to this Hamiltonian,
\begin{eqnarray*}
H = \vf \zuv.(\sgv\times\pv) + \gamma\sgz(k_x^2 + k_y^2)\label{eq:H}
\end{eqnarray*}
as suggested in Refs. [\onlinecite{Hong_12, Susskind_77}]. 
This k-space Hamiltonian is transformed to a real-space
Hamiltonian by replacing $k_x$ with differential operator 
$-i\frac{\partial}{\partial x}$, $k_x^2$ with $-\frac{\partial^2}{\partial x^2}$
and so on. The differential operators are then discretized in a square 
lattice using finite difference method to obtain the translational invariant,
real-space Hamiltonian, 
\begin{eqnarray}
\begin{split}
H = \sum_{i} c_i^\dagger \epsilon c_i 
+ \sum_{i} \left(c_{i,i}^\dagger t_x c_{i,i+1} + {\rm H. C.}\right)\\
+ \sum_{j} \left(c_{j,j}^\dagger t_y c_{j,j+1} + {\rm H. C.}\right)
\label{eq:H_r}
\end{split}
\end{eqnarray}
where $\epsilon = -4\hbar v_F\frac{\alpha}{a}\sgz$, 
$t_x = \hbar v_F\left[\frac{i}{2a}\sgy + \frac{\alpha}{a}\sgz\right]$,
$t_y = \hbar v_F\left[-\frac{i}{2a}\sgx + \frac{\alpha}{a}\sgz\right]$,
$a$ is the grid spacing and $\alpha \equiv \gamma a$ is a fitting parameter. 
For a grid spacing of $a=5$~\AA, the fitting parameter $\alpha = 1$ generates
a bandstructure that reproduces the ideal linear bandstructure
within a large energy window ($\sim\pm$0.5 eV) and gets rid of the Fermion 
doubling problem.
The discretized, real-space Hamiltonian given by Eq. \ref{eq:H_r} 
with parameters $\alpha = 1$ and $a=5$~\AA~is used for 
all of our NEGF calculations.

In order to calculate the charge and spin currents, we adopted the current 
density operator\cite{Zainuddin_11, Datta_97},
\begin{equation}
    I_{op} =  \frac{\i}{h} \big\{\Gn\Sigma^\dagger_m - \Sigma_m \Gn
 +  G\SigIn_m - \SigIn_m\dag{G} \big\}
\end{equation}
where $\Gn$ is the electron correlation function, $\Sigma_m$ is the self-energy
of contact $m\in \{S,D\}$ and $\SigIn_m$ is the in scattering matrix. The 
charge and spin currents are then given by,
$
    I_q(E) =  q {\rm Tr}\big\{I_{op}\big\}
$
and
$
    I_{\sv}(E) =  \frac{\hbar}{2} {\rm Tr}\big\{\sgv I_{op} \big\}
$
respectively. Since equilibrium spin current exists on the TI 
surface\cite{Hawthorn_spin_current_PR10,Tserkovnya_12}, this spin current includes
both equilibrium and non-equilibrium components. 
In order to obtain the non-equilibrium spin 
current, first we calculate equilibrium spin current, $I_{\sv_0}$ by setting 
$\mu_S = \mu_D = 0$ where, $\mu_S$ and $\mu_D$ are chemical potentials 
of the source and the drain contacts respectively. Then we calculate total 
(equilibrium + non-equilibrium) spin current $I_{\sv}$ by setting $\mu_D = 0$
and $\mu_S = qV_{DS}$. Finally, the total non-equilibrium spin current is 
obtained using $I_{\sv_{neq}}(E) =  I_{\sv}(E) -I_{\sv_0}(E)$ and integrating 
over all energies.

We found that the additional $\sgz$ term in Eq. \ref{eq:H_r} 
has an artifact. It gives a small non-zero $I_{sx}$ and 
$I_{sz}$ compared to zero values prediceted by our analytical model
using the exact Hamiltonian. However, this does not affect our conclusions 
since the focus is on $I_{sy}$.
Also, using the full 3D TI k.p Hamiltonian (descritized on 
the cubic lattice of a 3D TIPNJ slab) and the NEGF formalism, 
we verified that $I_{sx} = 0$ and $I_{sz} = 0$.
Although the full k.p Hamiltonian gives more accurate results for $I_{sx}$ 
and $I_{sz}$, it is computationally inefficient.

\section {Analytical Expression for Total Current}

The total current at energy $E$ is the sum of contribution from all electrons 
with positive group velocity along $\hat{x}$, 
$I(E) = W\sum_{v_x(\kv)>0}J(E,\theta)\delta(E - \ek)$
where $\delta$ is the Dirac delta function and $W$ is the width of the device. 
Replacing $\sum_{v_x(\kv)>0}$ with $\frac{A}{4\pi^2}\int_{v_x(\kv)>0} d^2k$ in 
this expression, using the delta function property 
$\delta(f(x)) = \frac{\delta(x - x_0)}{|f'(x_0)|}$
and integrating over all energy yield the general expression for total current,
\begin{equation}
I = \frac{W}{2\pi}\int dE  D(E) [f_S(E) - f_D(E)]
\int d\theta AJ(E,\theta)\label{eq:Iq1}
\end{equation}
where, $D(E)=\frac{1}{2\pi}\frac{|E+qV_n|}{\hbar^2\vf^2}$ is the density of 
states which has the units of eV$^{-1}$m$^{-2}$, and $f_S(E)$ and $f_D(E)$ are the Fermi-Dirac distributions of source and drain, respectively. 
Eqs. 2-4 (of main text) and Eq. \ref{eq:Iq1}  are valid for both symmetric and asymmetric 
built in potentials in $nn$, $pn$ and $pp$ regimes for all energies 
and hence can be used to calculate spin and charge current 
for large drain bias at room temperature. 
The solid lines of Figs. 2-3 were calculated using 
Eqs. 2-4 and Eq. \ref{eq:Iq1}. 

\section {The Angular Spin Current $\B{J_\w}$}

\begin{figure}[h]
    \includegraphics[width=3in]{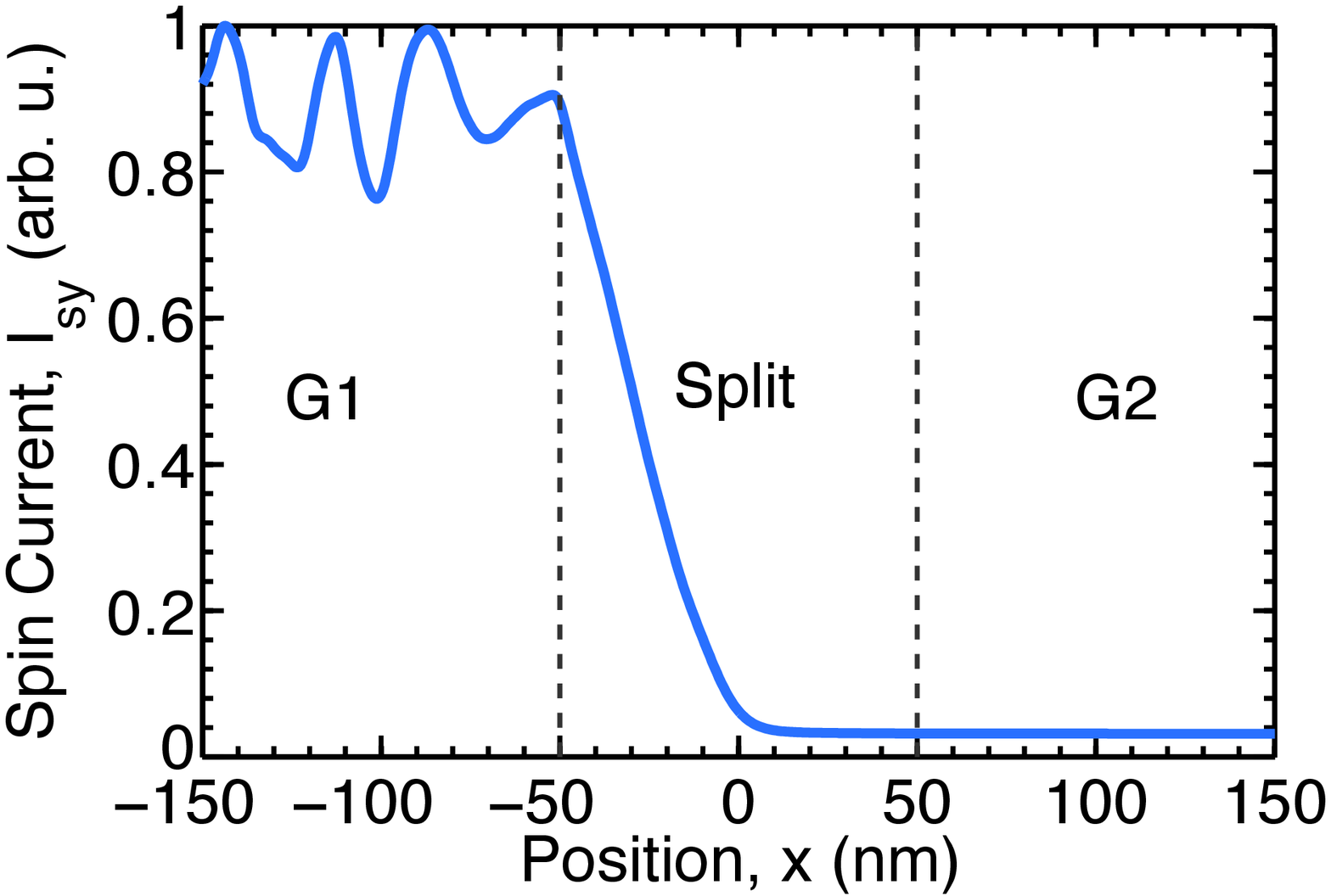}
    \caption{Spin current density as a function of position calculated using
    NEGF. Spin current is constant in the G2 ($p$) region. The small 
    oscillation in G1 ($n$) region is due to interference created by 
    reflected waves from the edges. The large change in the Split 
    region in due to rotation/precession of spin when electrons are reflected.
    In this simulation, we have used larger gates (100nm compared to 10nm in
    the original calculations) to illustrate the conservation of spin current
    in the uniform gate regions.} 
    \label{fig:Is_profile}
\end{figure}
It can be shown that the spin current operator that describes the angular 
motion/precession of spin in a 3D TI surface is given by
\begin{eqnarray}
\hat{\B{J}}_{\w} & = &  \hvf [-k_x\sgz\xuv - k_y\sgz\yuv + (k_x\sgx + k_y\sgy)\zuv].\label{eq:J_w}
\end{eqnarray}
The expectation value of $\hat{\B{J}}_{\w}$ for the TI surface eigenstate 
$    
\ket{\psi} = 1/\sqrt{2A}\left(1~~-si\e^{i \theta}\right)^{T}
    \e^{i\kv.\rv}
$ is $\B{J_\omega} = \B{0}$. Therefore, for a uniform TI channel where there 
is no scattering, the spin continuity equation becomes $\nabla.\B{J}_{\sv} = \B{0}$
in the steady state and the spin current is conserved.
More intuitively, in a uniform TI channel, the momentum of an electron 
does not change with time. Since the spin and momentum are locked,
the spin angular momentum also remains constant. Thus, there is no 
rotation/precession in spin and therefore, $\B{J_\omega} = \B{0}$
and the spin current is conserved.

Similarly, in the TIPNJ, the spin current is conserved inside the channel 
under the gates G1 and G2 where the potential profile is uniform. 
When an electron is reflected at the $pn$ junction interface, the direction 
of momentum changes by $\pi-2\theta_i$ accompanied by the same amount of 
change in the direction of spin angular momentum. In this case 
$\B{J_\omega} \neq \B{0}$ and the spin current is no longer conserved. 
The change in the spin angular momentum created by the reflection generates 
the spin torque.

This is also consistent with the spatial variation of spin current along 
the device calculated using NEGF as shown in Fig. \ref{fig:Is_profile}. 
Since there is no potential variation under the gates G1 and G2, there 
is no scattering and the spin current remains mostly conserved in these 
regions. The small oscillatory change in the G1 region is due 
to the interference created by the edge reflection in a finite-width 
device which is not included in the analytical model. The interference pattern 
can also be seen in the spin polarization plot in Fig. 2. On the other hand, 
inside the linear region, the electrons change direction which, in turn, 
results in a change in spin angular momentum. Therefore, spin current is not 
conserved. We also verified using NEGF that $I_{syS} - I_{syD} = I_{sy\omega}$ 
where, $I_{sy\omega} = \int J_{sy\omega} dS$, and $I_{syS}$ and $I_{syD}$ 
are the total spin currents at source and drain, respectively. Therefore, the 
difference between the spin currents at the source and the
drain terminal is the spin torque generated by the TIPNJ. 

In our analytical model, the effects of non-zero $\B{J}_\w$ inside the linear 
region are taken into account by constructing correct wave functions for the 
reflected and the transmitted waves. Given the correct wavefunctions, Eqs. (3) 
and (4) give correct spin currents everywhere for $x<-d/2$ and $x>d/2$ 
(neglecting the small oscillation due to edge reflections) since the spin 
current is conserved in these regions. In NEGF, the effects of 
$\B{J}_\w$ are taken into account automatically by the device Hamiltonian.

\bibliography{BIBLIOGRAPHY}

\end{document}